# 30 Years of Star Formation at UKIRT
*Chris Davis (JAC)*

It's safe to say that UKIRT's contribution to star formation at near-infrared (near-IR), mid-infrared (mid-IR) and even sub-millimetre (sub-mm) wavelengths has been considerable. From the early day's of single-detector photometers, through the development of 2-D arrays and complex multi-mode imager-spectrometers, to the present-day large-format imager WFCAM, UKIRT has offered the international community access to some of the world's most innovating, competitive, and versatile instrumentation possible. Suffice to say, UKIRT users have made the most of these instruments! Below I try to give a taste of the variety of star formation research that has come to pass at UKIRT (with apologies to those whose important work I fail to mention).

**Photometry of young stars – there's no hiding from UKIRT!**

Some of the earliest observations at UKIRT were conducted at mid-IR and sub-mm wavelengths, using single-channel bolometers such as UKT 7 and UKT 8 (affectionately known as Little Bertha and Big Bertha!), the popular and prolific UKT 14, and, for spectral line work, sub-mm receivers from Queen Mary College and the Rutherford Appleton Lab (e.g. Padman et al. 1985; White et al. 1986; Gear et al. 1988; Ward-Thompson et al. 1989). Moreover, almost from day one, UKIRT was open to observers who wished to bring their own instruments. For example, in one of the first papers to present mid-IR and far-IR photometry of embedded young stars, Davidson & Jaffe (1984) presented 400 μm photometry, obtained at UKIRT with the University of Chicago f/35 SMM photometer. The UKIRT data were used in conjunction with *Kaiper Airborne Observatory* photometry at shorter wavelengths to demonstrate the excess associated with cold circumstellar dust. At that time, similar observations existed for only one other object, B 335.

At shorter wavelengths, mapping the pre-main-sequence population really took off with the commissioning of UKIRT's first 2-D imaging array, IRCAM. Aspin, Sandell & Russell (1994) and Aspin & Sandell (1997) presented early photometry of dozens of young stars in NGC 1333 in Perseus, stressing (almost a decade before the launch of Spitzer) the importance of thermal imaging in their JHKL colour-colour and colour-magnitude diagrams, the longer-wavelength data being crucial for distinguishing the youngest sources from reddened background stars and weak-line T Tauri stars. Similar broad-band photometry was presented by Eiroa & Casali (1992) and Aspin & Barsony (1994), who used IRCAM to study the Serpens cluster and the red sources in LkHα 101, respectively. A few years later, Carpenter et al. (1997) analysed JHKL photometry of the Mon R2 cluster, complementing their IRCAM3 data of a 15'x15' region with CGS4 spectroscopy of 16 stars. Carpenter et al. found that two-thirds of the sources in the cluster exhibited infrared excess in the K and L-bands, and that the ratio of high to low mass stars was consistent with a Miller-Scalo Initial Mass Function (IMF).

In 1998 the UFTI commission team of Lucas and Roche used the new wide field and high spatial resolution afforded by this instrument to search for young stars and, particularly, for cool, low mass objects in and around the busy Trapezium cluster in Orion (Figure 1). They used IJH photometry to identify ~165 brown dwarf candidates, several of which were found to be young "free-floating planets" with masses below the deuterium burning limit (Lucas & Roche 2000). They followed up on these discoveries with deep spectroscopy at Gemini, identifying six sources with spectral types later than M9, consistent with planetary-mass objects (Lucas et al. 2006).

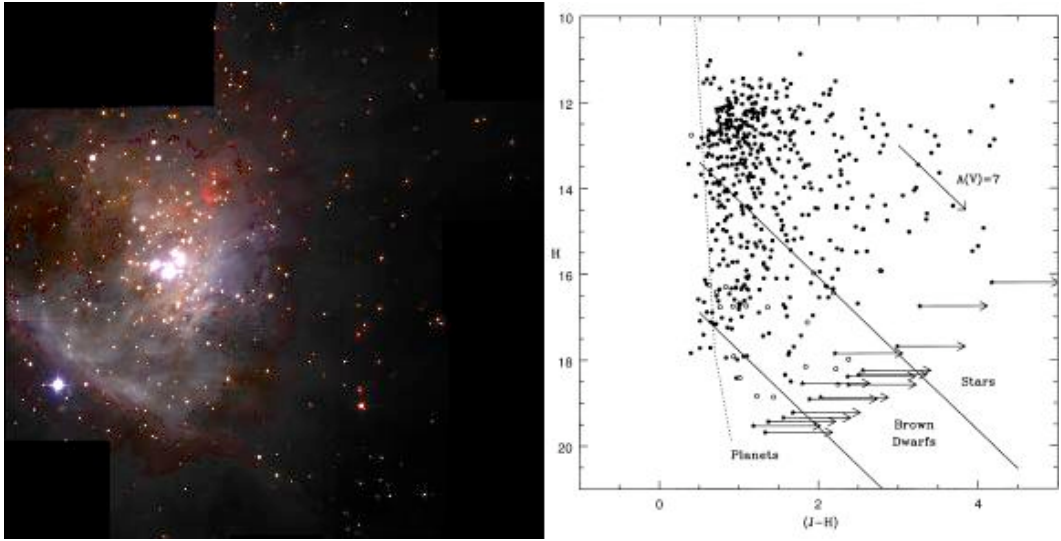

Figure 1: Left - IJH false-colour image of the Trapezium cluster in Orion. Right - colour-magnitude diagram extracted from these UFTI data, from which brown dwarfs and "free-floating planets" were identified. Data from Lucas & Roche (2000).

More recently, with the launch of the *Spitzer Space Telescope* and the availability of large-format near-IR cameras like WFCAM, there has been a surge in research on young stellar clusters. The UKIDSS Galactic Plane and Galactic Clusters surveys (described elsewhere in this volume) are yielding data that complement marvelously well the longer-wavelength *Spitzer* observations. These data will in the near future supercede 2MASS as the near-IR photometry of choice, because of their depth and superior resolution. In the meantime, PIs of non-UKIDSS projects are making excellent use of combined WFCAM and *Spitzer* datasets: for example, Kumar et al. (2007) have used photometry of some 60,000 stars to map the distributions of young, deeply embedded "Class I" sources and their more evolved contemporaries, the "Class II" T Tauri stars, across the massive star-forming region DR21/W75. They find the more abundant Class II sources to be spread widely throughout the region, the Class I proto-stars being tightly confined to regions of high extinction. Luhman et al (2008), working closer to home, have combined WFCAM and Spitzer data to search for discs around young brown dwarfs, while Wright & Drake (2008) have used WFCAM to search for near-IR counterparts to Chandra X-ray sources in the massive star-forming region CyG OB2. They identify counterparts for some 1500 sources.

WFCAM (and UKIRT as a whole) has also been popular with those wishing to conduct variability studies. Young stars -- FU Ori, EXor, T Tauri and Herbig Ae/Be stars -- are known to be variable on periods of months, weeks, or even days, due largely to bursts in accretion or to photospheric activity. Recently, Alves de Oliveira & Casali (2008) analysed WFCAM data, collected on 14 separate nights, to search for variable sources in the spectacular ρ Oph region (shown in Figure 2). They found 41% of the stellar population to be variable on periods of days and weeks; they associate this behavior with star spots and varying extinction. Lately, UKIRT has been monitoring a number of other regions on a nightly basis; currently (winter of 2009), monitoring of Orion is being conducted *in sync* with the *Spitzer Space Telescope*.

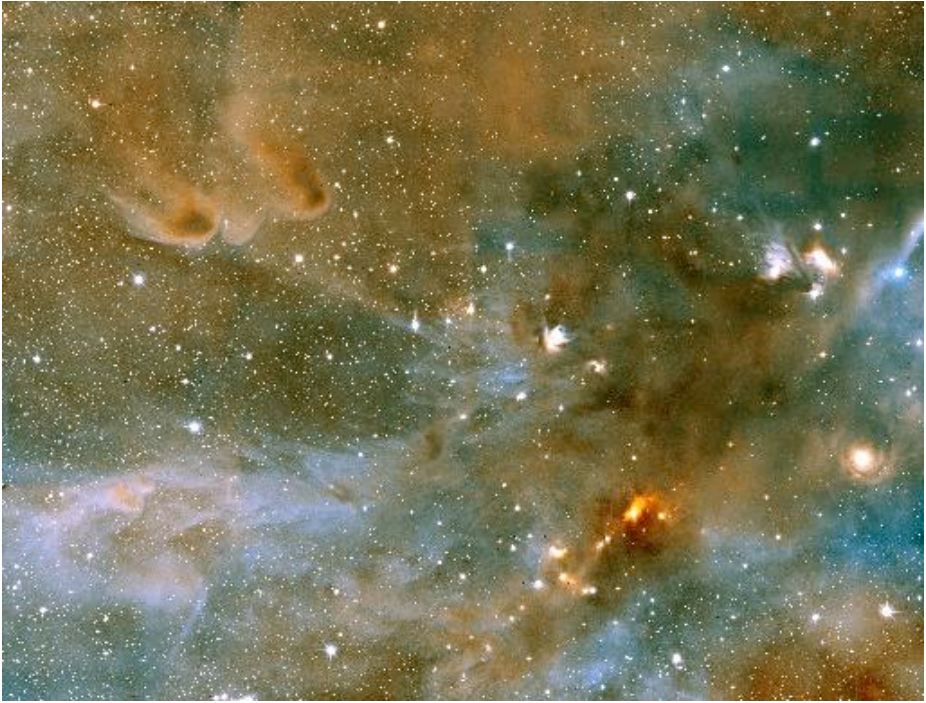

Figure 2: Deep WFCAM imaging of the low-mass star forming region Ophiuchus (Alves de Oliveira & Casali 2008).

**Excitation in the Interstellar Medium – some like it hot**

With its suite of imagers and spectrometers, UKIRT has played a pivotal role in advancing our understanding of the physical conditions associated with star forming regions and the Inter-Stellar Medium (ISM). Early observations of infrared nebula by Gatley et al. (1987), Hasegawa et al. (1987) and Tanaka et al. (1989 – see Figure 3) showed that the excitation of molecular gas in star forming regions was far from simple, being at best a combination of shock excitation and fluorescence. With the benefit of modern high-resolution imagers and spectrometers, this seems hardly surprising, given the complexity of star forming clusters (see e.g. the WFCAM image of the intermediate-mass star-forming region, AFGL 961, in Figure 3). But back in the day, with only a single-element detector to hand (UKT 9 in this case), a Circular Variable Filter (CVF) for order sorting, and a Fabry-Perot (FP) etalon for scanning in frequency to build up each spectrum pixel-by-pixel, these results were hard-won and certainly new and exciting.

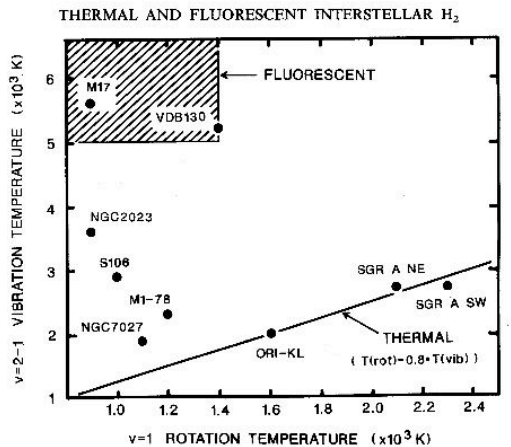 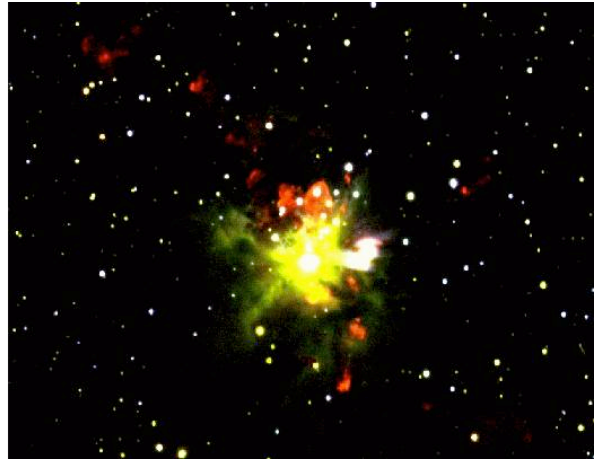

Figure 3: Left - Tanaka et al. (1989) used line ratios to identify the $H_2$ excitation mechanisms in a number of well-known galactic sources. Right - a more recent WFCAM JHH$_2$ image of the intermediate-mass star-forming cluster AFGL 961, showing the complexity of the region.

In the '80s and early '90s, UKIRT very much led the way in studies of galactic nebula (Jourdain de Muizon et al. 1986; Gatley et al. 1987; Brand et al. 1988; Burton et al. 1989; Geballe et al. 1989; Chrysostomou et al. 1993). For example, Burton et al. (1990) observed $H_2$ 1-0S(1) line profiles at high spectral resolution in a number of galactic nebulae, finding the lines to be narrow and peaked at ambient velocities – a result they interpreted in terms of non-thermal excitation (fluorescence). At about the same time, Geballe & Garden (1990), building on earlier UKIRT observations (Geballe & Garden 1987) were mapping OMC-1 (in Orion) in pure-rotation $H_2$ 0-0S(9) and CO 1-0P(8) emission at 4.7 μm; in this region at least, much broader lines were observed, which were symptomatic of shock excitation. As we shall see in the next section, work pushed on in this area largely via studies of individual low-mass objects, as spatial resolution improved and observers gained access to instruments equipped with 2-D arrays of pixels.

**Accretion and outflow – what goes down, must come up!**

Access to a wonderful new cooled grating 1-5 μm spectrometer, CGS4, drove observational studies of perhaps the most fundamental processes associated with star formation – accretion and outflow. As part of her thesis work in Edinburgh, Chandler, together with her collaborators, obtained some of the first high-resolution spectroscopic observations of ro-vibrational CO band-head emission at 2.3 μm (Chandler et al. 1990 – see also Figure 4). In a sample of a half-dozen young stars they observed a variety of band-head profile shapes, which they modeled in terms of excitation in the inner regions of a rotating Keplarian disc (see also Chandler et al. 1995). Similar observations were later obtained by Casali & Matthews (1992), who observed CO in absorption; by Aspin (1994), who detected CO band-head emission scattered by the dense gas that envelopes the massive young star GGD-27; and by Reipurth & Aspin (1997), who surveyed a number of outflow sources in the K-band at low spectral resolution, including some well-known FU Ori variables (all of which possessed strong CO in absorption).

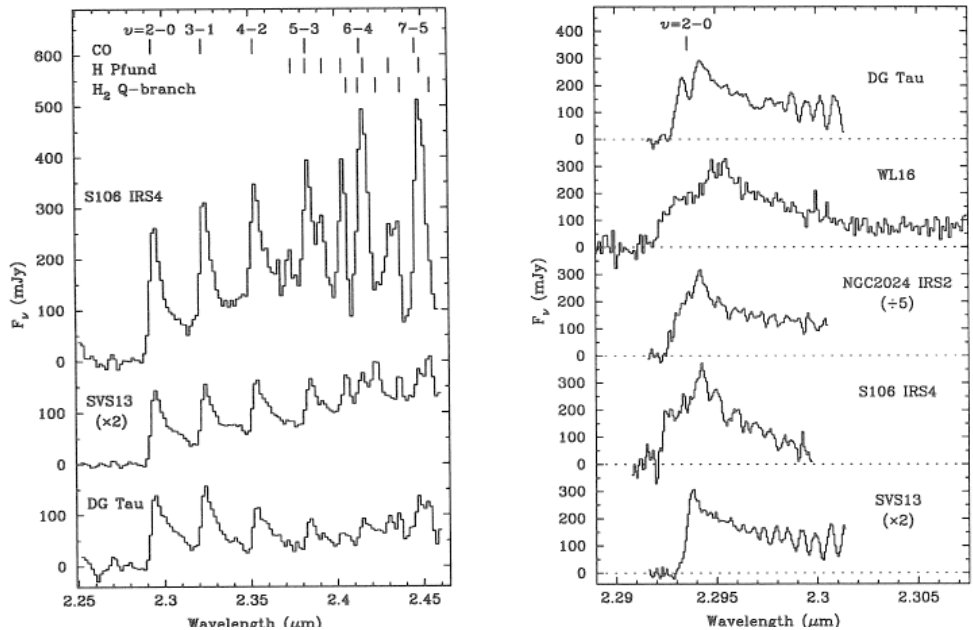

Figure 4: CGS4 spectroscopy of accreting proto-stars, at low spectral resolution (left) and at high spectral resolution using the echelle (right). The CO band-heads in these spectra were thought to be associated with the warm, dense inner regions of accretion discs (Chandler et al. 1990).

Focusing on more evolved sources, Folha & Emerson (2001) used CGS4 in an ambitious survey of 50 T Tauri stars. Observing with the echelle, they used Paβ and Brγ hydrogen recombination lines as probes of accretion and outflow processes. They observed a variety of line shapes including P Cygni profiles which they interpreted in terms of magnetic accretion with velocities of hundreds of kilometers per second.

A few years later, Sheret, Ramsay Howat & Dent (2003), inspired by claims in the literature of detections of pure-rotational $H_2$ emission from the discs of pre-main sequence stars – discs viewed as perhaps the precursors to proto-planets – tried to confirm these observations with Michelle (during its brief stay at UKIRT). They obtained a marginal detection of the 4-2 emission line a 12.2 μm in one of their two targets (AB Aur), though failed to detect any emission from the other.

Studies of outflows at UKIRT were very much inspired by the early work of Zealey, Williams and collaborators (Zealey et al. 1984, 1992), who used IRCAM to image a number of well-known Herbig-Haro (HH) objects in $H_2$ 1-0S(1) line emission (Figure 5), and by Zinnecker et al. (1989), who used IRCAM with the FP to resolve $H_2$ line profiles in a number of HH objects. Carr (1993) likewise used IRCAM with the FP, though rather than observe multiple sources, he mapped $H_2$ line profiles across one of the brightest HH objects known at that time, HH 7 (Figure 6); similar observations, of the explosive outflow activity in OMC-1 in Orion, were conducted by Chrysostomou et al. (1997). CGS4 was later used to map $H_2$ profiles across the L 1448 and DR 21 outflows (now known as MHO 539 and MHO 898/899, respectively), and along two of the spectacular "bullets" that emanate from the Orion nebula (Davis et al. 1996a, 1996b; Tedds, Brand & Burton 1999). At about the same time Fernandez and co-workers used CGS4 with its low-resolution grating to examine gas excitation in a number of flows (Figure 6); they found that in HH 7 and DR 21 a combination of shocks and fluorescence was needed to account for their excitation diagrams (Fernandez & Brand 1995; Fernandez, Brand & Burton 1997).

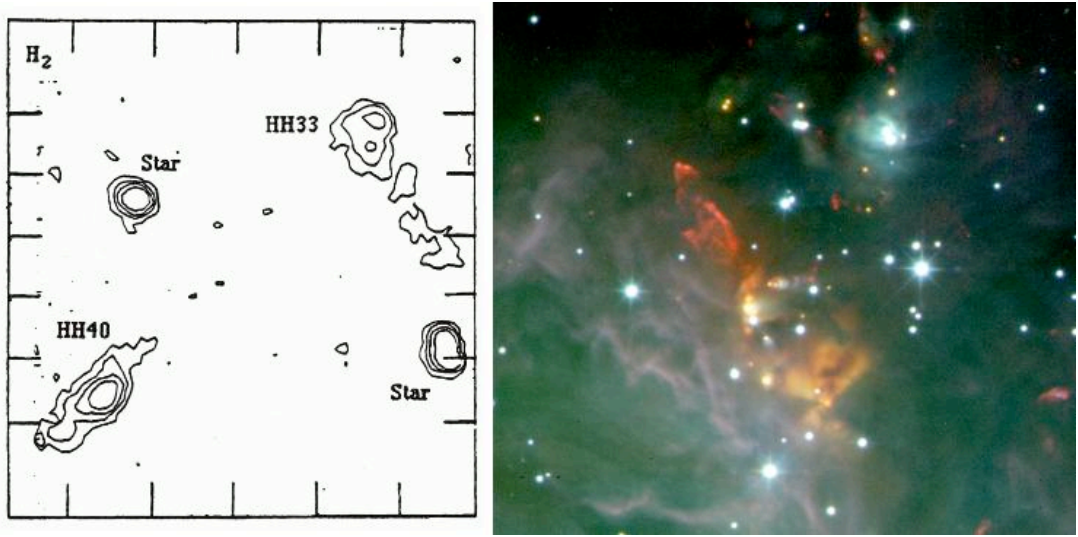

Figure 5: Left - early $H_2$ imaging of outflows from young stars, in this case the Herbig-Haro objects HH 33/40 (Zealey et al. 1992). Right - WFCAM JK$H_2$ false-colour image of jets and embedded young stars in the OMC-2 region of Orion (Davis et al. 2009).

More recently, Davis et al. (2003) and Whelan et al. (2004) have used CGS4, its echelle grating, and a technique known as spectro-astrometry to measure the relative positions of $H_2$, [FeII] and Pa$\beta$ emission-line peaks to within a few tens of AU of a number of outflow sources; these observations help constrain jet collimation and acceleration models. The Integral Field Unit (IFU) in UIST has also been used to image, in various emission lines, collimated "micro-jets" from a number of intermediate-mass young stars (Figure 7; Davis et al. 2004, Varricatt et al., in prep.)

We've come a long way since the days of single-object imaging and spectroscopy in the mid-'90s: earlier this year, Davis et al. (2009) mapped 8 square degrees in Orion with WFCAM, identifying well over a hundred $H_2$ flows (see e.g. Figure 5), measuring the proper motions of multiple knots in 33 of them, and associating the vast majority with embedded (Spitzer) proto-stars and dusty proto-stellar cores. With the completion of the UKIDSS $H_2$ survey of the Taurus-Auriga-Perseus star-forming complex, and the recent approval of PATT time for the UWISH2 narrow-band survey of the galactic plane (http://astro.kent.ac.uk/uwish2), the future looks bright for outflow studies at UKIRT.

**Massive star formation and UKIRT's love affair with DR 21**

Over the years, one of the most popular targets at UKIRT has been the spectacular massive star forming region, bright radio source, nest of masers and H II regions, and complex of molecular outflows collectively known as DR 21. As with many areas in astronomy at UKIRT, the earliest observations were conducted at sub-mm wavelengths. For example, Richardson et al. (1986) produced a very JCMT-*esque* study of DR 21 using receivers from Queen Mary College. They published a remarkably detailed work that included CO, CS, HCN, HCO+, $H^{13}CO+$ and CS spectroscopy, as well as continuum observations at 20 μm and 300 μm (Figure 8). With these data they were able to map the distribution of high velocity molecular gas around the central HII regions, and model the overall ambient gas distribution.

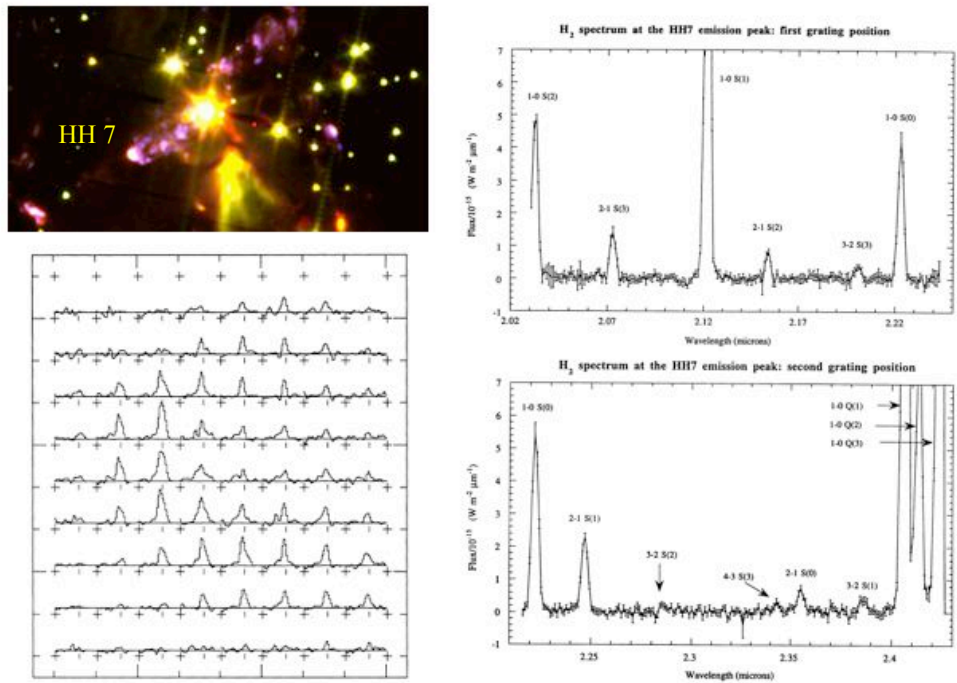

Figure 6: High-resolution IRCAM+FP (left) and low-resolution CGS4 (right) $H_2$ spectroscopy of the HH 7 bow shock (Carr 1993; Fernandez & Brand 1995). HH 7 is evident to the south-east of the bright young star SVS13 in the false-colour WFCAM/Spitzer image shown top-left.

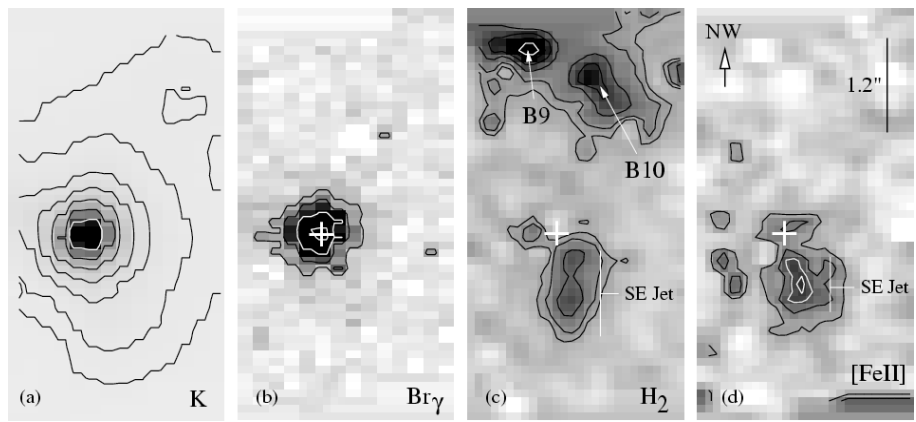

Figure 7: UIST IFU spectral images showing the collimated jet associated with the luminous young star IRAS 18151+1208 (Davis et al. 2004). The jet is clearly seen in continuum-subtracted

H$_2$ and [FeII] images; the Brγ coincident with the source is thought to be associated with magnetospheric accretion flows.

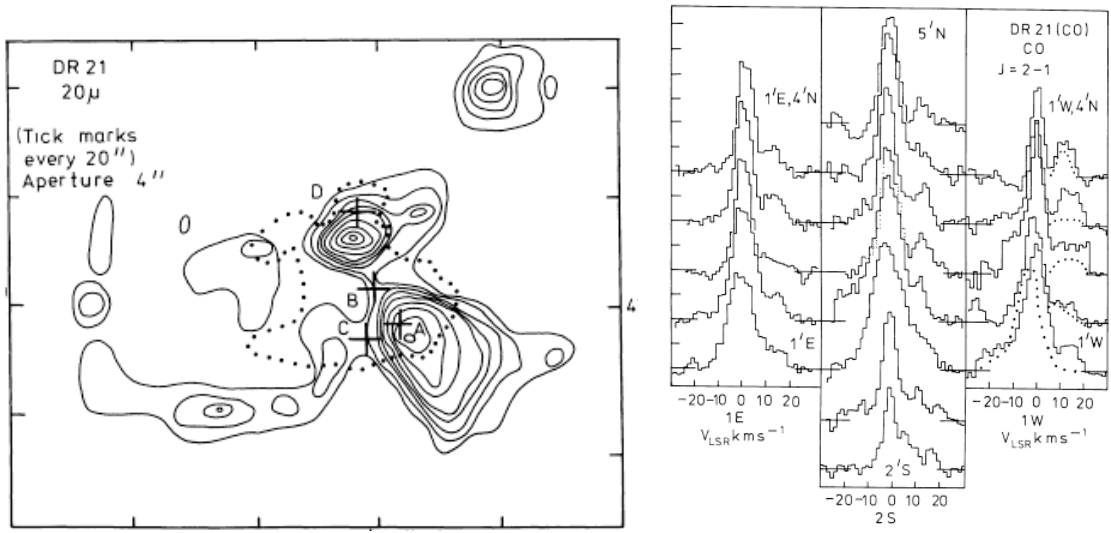

Figure 8: Mid-IR and sub-mm observations of the massive star forming region/cluster of compact HII regions known as DR 21; left – a 20 μm map taken with a 4" aperture, right – CO J=2-1 spectra. All data are from Richardson et al. (1986).

At shorter wavelengths, Roelfsema, Goss & Geballe (1989) used CGS2 and the FP to probe the ionized gas in DR 21 via Brα observations at high resolution, mapping the emission and adjacent 4 μm continuum across the central region. Meanwhile, Garden and co-workers were focusing on the bright bipolar outflow associated with DR 21, first mapping the H$_2$ emission, one velocity-resolved spectrum at a time with a CVF and the FP (Garden et al. 1986). These were pain-staking observations that involved switching the FP between line and line-free frequencies every 3 seconds, and sampling blank sky every 10 pointings. The final map of DR 21 comprised observations at 400 positions. Garden et al. followed up on this work with IRCAM imaging and high resolution H$_2$ line profile mapping (Garden et al. 1990, 1991).

These data (and similar studies of Orion, e.g. Brand et al. 1988) subsequently inspired a slew of theoretical papers aimed at interpreting line shapes and excitation diagrams in terms of planar and curved (bow-shaped) "Jump" shocks and magnetically-cushioned "Continuous" shocks (Smith & Brand 1990a, 1990b, 1990c; Smith, Brand & Moorehouse 1991a, 1991b, Smith 1994). These tried-and-tested models still represent some of the most comprehensive work done on molecular shock physics.

DR 21 was of course not the only high mass star-forming region targeted by UKIRT observers. Bunn, Hoare & Drew (1995) used Brα, Brγ and Pfγ as tracers of high velocity winds in a number of massive young stars (see also Drew, Bunn & Hoare 1993; Lumsden & Hoare 1996, 1999) and, in a very ambitious new project, Varricatt et al. (2010) have recently used UFTI to search for collimated molecular outflows in 50 massive star forming regions. In regions where collimated jets have been observed, follow-up CGS4 echelle and UIST IFU observations have also been secured.

**Polarimetry of star forming regions**

Polarimetric observations have for many years been a main-stay at UKIRT (see the article by Jim Hough in this volume), and a review of star formation would not be complete without mentioning some of the innovative observations and ground-breaking results obtained at UKIRT. Polarimetry has been possible largely because of the continued support from the University of Hertfordshire; users have also benefited from improved instrument design (putting the Wollaston prism downstream of the slit rather than upstream, as was the case with CGS4), enhanced data acquisition, and reliable imaging and spectro-polarimetry data reduction software (as part of the ORAC-DR program). In fact, in the last few years polarimetry has become so straight-forward, particularly with UIST, that during recent Cassegrain blocks it was incorporated into queue scheduling and the service observing program.

Linear spectro-polarimetric observations of scattered light from dust grains around T Tauri stars were made as early as November 1979 using the Hatfield Polarimeter (Hough et al. 1981). A few years later $H_2$ and K-band imaging polarimetry of the Orion BN-KL nebula were made with the Kyoto-UKIRT infrared polarimeter and UKT 9 (Hough et al. 1986); these observations revealed for the first time a molecular hydrogen reflection nebula around the flow from BN, and represented one of a number of early collaborations between UK and Japanese astronomers at UKIRT (see also Nagata et al. 1986; Yamashita et al. 1987, etc.; note also that Chrysostomou et al. [1994] subsequently used $H_2$ line polarisation measurements to map a twist in the magnetic field about IRc2, the most luminous young star in the region). Sato et al. (1988) observed 20 sources in Ophiuchus in polarized light, thereby mapping the degree of polarisation and changes in the polarisation position angle across the densest regions of the cloud, while Aitken et al. (1988) used the UCL array spectrometer with a wire grid to observe polarisation at 10 and 20 μm in AFGL 2591, thereby showing how spectro-polarimetry could be used as a sensitive indicator of grain chemistry.

In the early '90s polarimetrists were not slow to take advantage of the 2-D arrays cameras arriving at UKIRT. In a series of papers, Minchin et al. (1991a, 1991b, 1991c) mapped and modeled the polarised reflection nebulae associated with a number of young IR sources, while Whittet et al. (1992) used UKIRT and the Anglo-Australian Telescope to examine the wavelength dependence of polarisation, developing at the same time a comprehensive catalogue of polarised standards that is still used today.

In recent years the full suite of Cassegrain instruments at UKIRT has been used for polarimetry: Chrysostomou et al. (2000) used IRCAM3 to map the circularly polarised emission across OMC-1 (Buschermohle et al. [2005] conducted similar observations with UFTI); Kuhn et al. (2001) instead used IRCAM3 to search for circumstellar discs around young stars; Holloway et al. (2002) measured the polarisation of the 3 μm water-ice feature towards a number of young stars; Oudmeijer, Drew & Vink (2005) observed polarised line emission from massive young stars with UIST in spectroscopy mode, while

Hales et al (2006) used UIST in imaging mode to map the polarised light around a number of dusty low-mass young stars. Hough et al. (2008) recently used UIST to examine grain alignment through spectro-polarimetry of the 4.7 μm CO ice feature, and very recently Wisniewski and co-workers have been using UIST in its newest mode, coronographic-imaging-polarimetry, to trace the scattered light discs around very bright Herbig Ae stars (Figure 10).

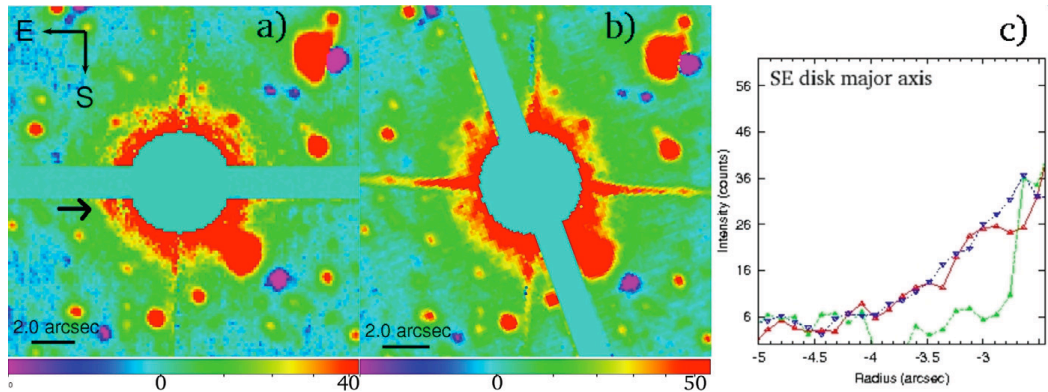

Figure 10: Coronagraphic-imaging-polarimetry of the Herbig Ae proto-planetary disc system HD 163296 (Wisniewski et al., in prep.). Panels a) and b) show the bright star at two UIST position angles after subtraction of a PSF star. Note that the features associated with the scattered light disc remain at both position angles. Panel c) presents radial profile cross-sections taken across the south-east quadrant at each UIST position angle (red and blue lines). The green line shows a similar profile for a source with no known disc. Note that data were taken through a Wollaston prism and waveplate; analysis of the polarised light associated with the disc is pending.

**The Future**

Clearly, UKIRT has made a major contribution to the field of star formation, through observations at near-IR, but also mid-IR and sub-mm wavelengths. British astronomers and their collaborators have always been quick to take advantage of the variety of modes made available to them, and continue to do so in the era of wide-field astronomy and large, multi-national collaborations on legacy surveys like UKIDSS. WFCAM observations are hugely complementary to *Spitzer* mid-IR imaging and photometry, JCMT/HARP sub-mm molecular line observations of large-scale cloud structure and dynamics, and JCMT/SCUBA-2 mapping of the cold dust emission in dense pre-stellar and proto-stellar cores. It is in these areas that UKIRT, as she passes her 30$^{th}$ birthday, is now making an impact on star-formation research, and will continue to do so for a number of years to come.

*12 November 2009*